# Acoustic Monitoring of Inelastic Compaction in Porous Granular Materials

Vincent Canel[1,2], Xiaoping Jia[1*], Michel Campillo[2], and Ioan Ionescu[3]

[1]*Institut Langevin, ESPCI Paris, Université PSL, CNRS, Paris 75005, France*
[2] *ISTerre, CNRS, UMR 5275, Université Grenoble Alpes, 38000 Grenoble, France*
[3]*LSPM, CNRS UPR 3407, Université Paris 13, 93430 Villetaneuse, France*

## ABSTRACT

We study the transition from cohesive to non-cohesive granular states of synthetic rocks under oedometric loading, combining simultaneous measurements of ultrasound velocity and acoustic emissions. Our samples are agglomerates made of glass beads bonded with a few percent of cement, either ductile or brittle. These cemented granular samples exhibit an inelastic compaction beyond certain axial stresses likely due to the formation of compaction bands, which is accompanied by a significant decrease of compressional wave velocity. Upon subsequent cyclic unloading-reloading with constant consolidation stress, we found the mechanical and acoustic responses similar to those in non-cohesive granular materials, which can be interpreted within the effective medium theory based on the Digby's bonding model. Moreover, this model allows P-wave velocity measured at vanishing pressure to be interpreted as an indicator of the debonding on the scale of grain contact. During the inelastic compaction, stick-slip like stress drops were observed in brittle cement-bonded granular samples accompanied by the instantaneous decrease of the P-wave velocity and acoustic emissions which display an Omori-like law for foreshocks, i.e., precursors. By contrast, mechanical responses of ductile cement-bonded granular samples are smooth (without visible stick-slip like stress drops) and mostly aseismic. By applying a cyclic loading-unloading with increasing consolidation stress, we observed a Kaiser-like memory effect in the brittle cement-bonded sample in the weakly damaged state which tends to disappear when the bonds are mostly broken in the non-cohesive granular state after large-amplitude loading. Our study shows that the macroscopic ductile and brittle behavior of cemented granular media is controlled by the local processes on the scale of the bonds between grains.

## I. INTRODUCTION

Deformation of rocks involves damage process such as microcracks nucleation and propagation. During the deformation process, the damage localization can lead to the creation of macroscopic fractures (e.g., shear bands) and to the failure of the material associated with a dramatic decrease of its strength and modulus. On the geological scale, earthquakes or fault core sliding occur naturally in response to long-term deformation produced by plate tectonics. However, the way the damage and the cohesiveness of the damaged rocks control the frictional slip and the seismic patterns is not well understood [1]. Indeed, typical faults consist of a narrow fault core of almost purely granular material, where earthquake slip localizes [2], surrounded by a fractured crust (off-fault damage zone) whose fracture density decreases with distance away from the fault core [3]. This damage zone exhibits a unique mechanical and hydraulic behavior which does affect earthquake ruptures. In turn, the earthquake ruptures impart a wide range of loading rates on the surrounding damage zone which affect the hydro-mechanical properties of the damage zone [4].

---
* xiaoping.jia@espci.fr

A similar behavior can be reproduced in the laboratory via triaxial loading experiments on initially fractured rocks (e.g., containing a sawcut), a likely situation in a pre-existing seismic fault [5]. According to observations, stable sliding or stick-slip instabilities depend on loading conditions, well rationalized by the rate and state friction law [6, 7]. However, for initially unfractured rocks, two distinct mechanical responses and failure modes are observed [8]. For brittle materials, the failure occurs due to the cooperative interaction of local (micro-)ruptures, i.e., damage localized into a narrow shear band (inclined at 30°-45° compared to the compressive load direction) as loading proceeds, that is experimentally observed by acoustic emission (AE) source localization (cracking noise). Ductile behavior associated with diffuse damage (macroscopic plasticity) can also be observed for the same materials by increasing the confining pressure. In both cases, the acoustic emission (AE) event distributions exhibit power law behavior pointing to long range correlations; nevertheless, their behaviors appear to be akin respectively to the first-order and critical phase transitions [1, 9]. These observations are confirmed by various numerical models and simulations [10-12].

In addition to the formation of shear bands, compaction bands have also been observed in porous (granular) sandstones at a critical confining (hydrostatic) pressure, monitored by the AE source location detections and the elastic wave velocities evolution. They are localized compressive deformation zone mostly perpendicular to the main stress direction due to grain crushing and pore collapse [13-15]. If the increase of local density under load results in strengthening, formation of a compaction band may not preclude later formation of a shear band or failure [16]. Such inelastic compaction and associated porosity reduction play an important role in the diagenesis of sandstones and may affect the reservoir stability during hydrocarbon production.

In this work, we focus our investigation on the damage process during inelastic compaction. To this end, we perform controlled laboratory experiments on granular porous materials under oedometric loading for which shear banding is likely excluded. Our synthetic rocks are made of glass beads bonded with a few percent of cement, either ductile or brittle. Compared to other previous works [17, 18] here we monitor acoustically the damage and fracture process by measuring both the change in compressional wave velocity (active detection) [19] and the acoustic emission (passive detection) if any [20]. Our aim is twofold. First, unlike porous sandstones, the inelastic compaction considered here shall be associated with the debonding instead of grain crushing. These cemented granular materials allow a close comparison between experiments and theory based on the Digby granular model. By measuring the compressional wave velocity at vanishing confining pressure, this model provides a qualitative indicator of the bond damage-breakage induced by cyclic loading-unloading from (continuum) cemented state to (discrete) granular state [21]. Secondly, by controlling the local interaction and damage process through ductile-cement and brittle-cement bonded glass beads, we seek for the possible link with the associated ductile and brittle behaviors on the macroscopic scale.

In the following sections, we first describe the preparation protocol of cemented granular samples (section II.A) and combined oedometric tests, ultrasonic measurements and acoustic emission detections in these materials, under constant (section II.B) and increasing (section II.C) consolidation stresses, respectively. In Section III, we analyze the compressional wave velocity and damage process within the effective medium framework based on a heuristic bonded contact (sections III.A and III.B). We also discuss the Omori-like behavior and the memory effect (Kaiser effect) with acoustic emission analysis at constant and increasing consolidation stresses (section III.C).

## II. EXPERIMENTS

### A. Sample preparation and oedometric loading

Our cemented granular materials are composed of polydisperse glass beads of diameter $d$ = 400-800 μm bonded with either ductile or brittle cement. The packing fraction of beads is $\phi_s \approx 60\%$ and the cement volume fraction $\phi_c \approx 1.5\%$ (compared to the total volume of the samples). One of the cylindrical samples is shown in Fig. 1(a), with diameter $D$ = 32 mm and thickness $e_0$ = 8.0 mm. The ductile cement is made of eicosane that tolerates large deformation before breakage, while the brittle cement is salol that breaks readily under stress. The cement powders are added to glass beads in a cylindrical mould and then placed into an oven at a temperature just greater than the melting point of the cement (around 40°C). Once the cement is melted after a few minutes of heating, we stir the mixture to distribute evenly the cement with the beads. The mixture is then closed in the mould and submitted to cycles of oedometric loadings and unloading that allow to compact the sample before the cooling and solidification of the cement at the room temperature.

Once cooled, cemented granular samples are placed into a cylindrical oedometric cell of the same diameter. Two large longitudinal broadband transducers (of diameter $D$ = 32 mm) are put in contact with the sample as pistons (in green, Fig. 1(b)). They transmit the load to the sample from a uniaxial electromechanical press, under a controlled displacement speed (i.e., axial strain-driven), monotonic or cyclic. For the range of axial loading force $F$ applied here, we have verified that the beads remain intact (not crushed) after the test. Acoustic measurements are performed during the external loading to monitor the damage process of the sample under compression.

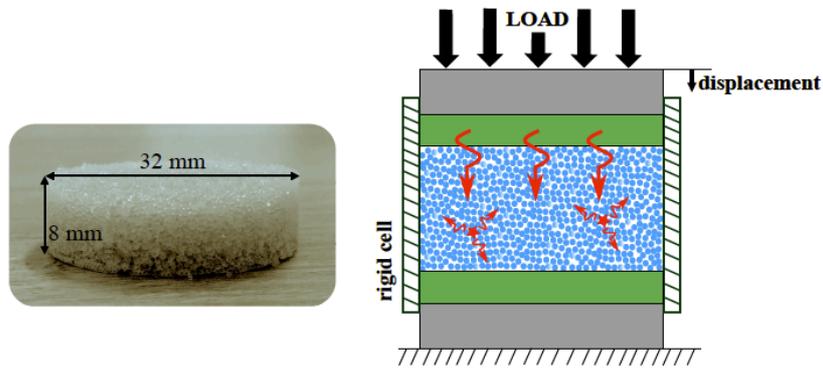

Figure 1. The cemented granular material (a) is placed between two ultrasonic transducers (sketched in green) in a rigid cell closed by two pistons, i.e., oedometric testing (b). The upper piston transmits the load through a controlled displacement (axial strain-driven). The active acoustic probing is conducted with the source ultrasonic transducer (top) transmitting a short pulse (long red arrows) and the detector (bottom), while the passive detection, i.e., acoustic emissions (red stars) are recorded by the ultrasonic detector.

### B. Damage evolution under cyclic loading with constant consolidation stress

*Mechanical response*: Fig. 2 depicts the loading protocol where the loading speed is imposed at 100 μm/s (strain-driven), and both displacement and force are measured as a function of the test time in the ductile- and brittle-cement bonded granular samples, respectively. When the load reaches the maximum amplitude $F = 3000$ N (corresponding to an axial stress of $P = F/(\pi D^2) \approx 3.7$ MPa), the sample is unloaded to $F = 60$ N ($P \approx 75$ kPa) and after then submitted a dozen cycles of unloading-loading between 60 N and 3000 N. A non-zero unload lower limit was used to keep the contact between transducers and the granular sample for ensuring reproducible measurements.

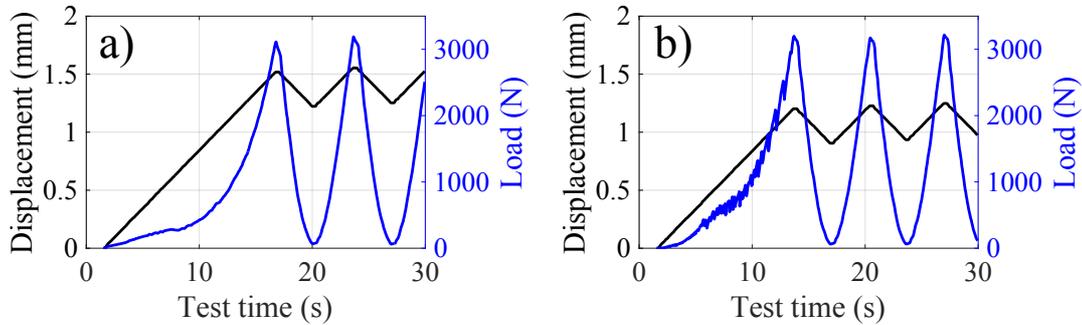

Figure 2. Both displacement and force are measured as a function of the test time at an imposed loading speed (strain-driven) with the ductile- (a) and brittle- (b) cement bonded granular samples.

Fig. 3 shows the cross-plots of the loading force as a function of the displacement in the ductile- and brittle-cement bonded granular samples. For the ductile-cement bonded sample (Fig. 3(a)), the load first increases linearly with the displacement from A to B, then deviates nonlinearly from B to E, pointing to plastic deformations. More specifically, we observed the appearance of a plateau in the load-displacement curve from C to D for numerous tests with different cemented samples. It suggests likely the growth of microscopic cracks (initiated at the point B) into mesoscopic or macroscopic fractures in cemented samples, leading to a plastic deformation on the macroscopic or sample scale. Mechanical responses are fundamentally different between the first loading and subsequent unloading-reloading cycles. The latter is reminiscent of the behaviour observed in non-cohesive, dry granular materials (unconsolidated) where small hysteretic loops are detected between the loading and the unloading paths. They are accompanied by a slight compaction [22] and tend to an almost stationary narrow loop, displaying the hertzian-like force-displacement relationship [23].

For the brittle-cement bonded granular sample (Fig. 3(b)), we observe the similar general behaviour during the first monotonic loading as above. However significant intermittent fluctuations corresponding to stress drops are present, which make difficult the identification of the previous plateau from C to D visible in Fig. 3(a) (see discussion below). These stress drops highlight transient stick-slip behavior, which may be caused by fracture nucleation or shear banding inside the loaded sample. Nevertheless, such stress drops are no longer observed during the subsequent unloading-reloading cycles (between E and F) in which a dry granular medium behaviour is recovered, as for the ductile cement bonded sample.

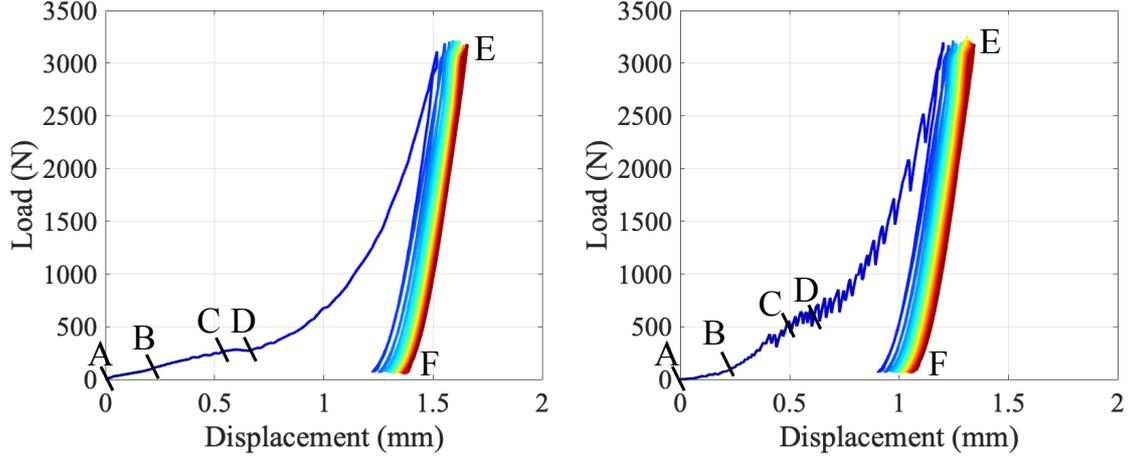

Figure 3. Loads measured as functions of displacements (strain-driven) during the oedometric testing with ductile (a) and brittle (b) cement bonded granular samples, respectively. The response is remarkably different between the first loading (from A to E) and subsequent unloading-reloading (between E and F).

*Ultrasound velocity measurement*: Our *active* acoustic probing relies on the investigation of one sinusoidal pulse propagation through the sample during the mechanical test. This small-amplitude wave is centered at $f = 100$ kHz with a corresponding wavelength about $\lambda \sim 10\text{-}20$ mm for a compressional wave velocity measured as $V_P \sim 1000$ m/s and 2000 m/s in dry non-cohesive and cemented granular media, respectively [17, 23]. In this frequency range, the wavelength is much larger than the bead size $d$ ($\lambda \gg d$) and therefore coherent waves are dominant compared to scattered waves. The detection of these coherent waves is also enhanced by the large transducer of diameter $D$ ($\gg d$) [24]. Typical transmitted acoustic signals during the oedometric test in a ductile-cement bonded sample are shown in Fig. 4. The test time ($\sim 30$ s) refers to the mechanical loading where the indication A to F corresponds to different loading steps defined in the previous section, while the acoustic time refers to the wave propagation time ($\sim 30$ μs). The wave propagation is much faster than the quasistatic mechanical loading and thus provides a snapshot of the sample elastic properties at different loading stages (mechanical relaxation of the sample is ignored here).

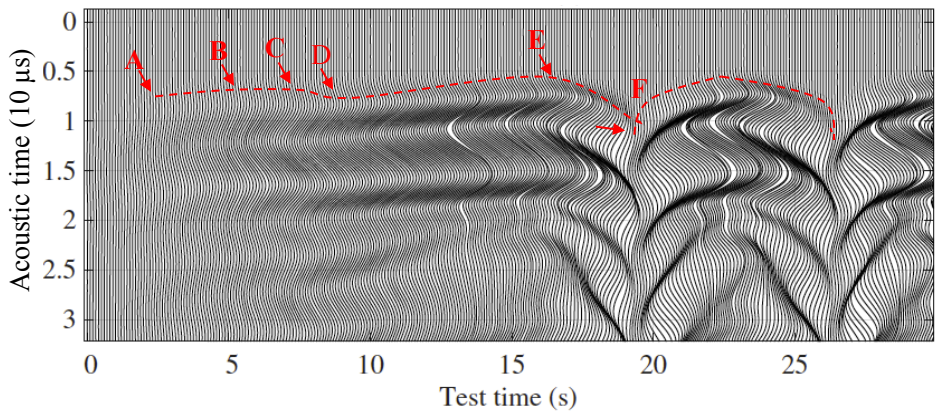

Figure 4. Transmitted acoustic signals during the oedometric loading in a ductile-cement bonded sample. The red line indicates the arrivals of the direct P-waves (when 5% of the 1st peak amplitude is reached) and the letters refer to different loading steps indicated in Fig. 3(a).

More specifically, we characterize the material state or properties such as damage via the compressional coherent wave (also called P-waves) velocity $V_P$, measured by the time of flight of the signal picked at 5% of the 1$^{st}$ peak amplitude. The waveform is indeed not only shifted in time because of the velocity change, but also deformed since its spectral content evolves with the state of the material due to the change in wave attenuation and scattering. This early time of flight is less influenced by the waveform distortion, allowing us to neglect such effect as a first approximation.

The evolutions of the P-wave velocities are shown in Fig. 5 for the ductile- and the brittle-cement bonded granular samples. These velocities are greater than those in a non-cohesive granular medium at the first loading for a comparable confining pressure (Jia et al, 1999) due to the bonds that enhance the P-wave velocity [17]. During the monotonic loading from A to E, the P-wave velocity in both cemented samples exhibit a complex evolution with the load or the controlled displacement. More precisely, $V_P$ first increases (A-B), decreases slightly (B-C) then drastically (C-D) and eventually slightly increases (D-E). The following cycles (E-F) are described by the hysteretic loops accompanied with slight compaction, pointing to the behaviour of a non-cohesive frictional granular medium [24] where $V_P$ scales on confining pressure as $V_P \sim P^{1/6}$.

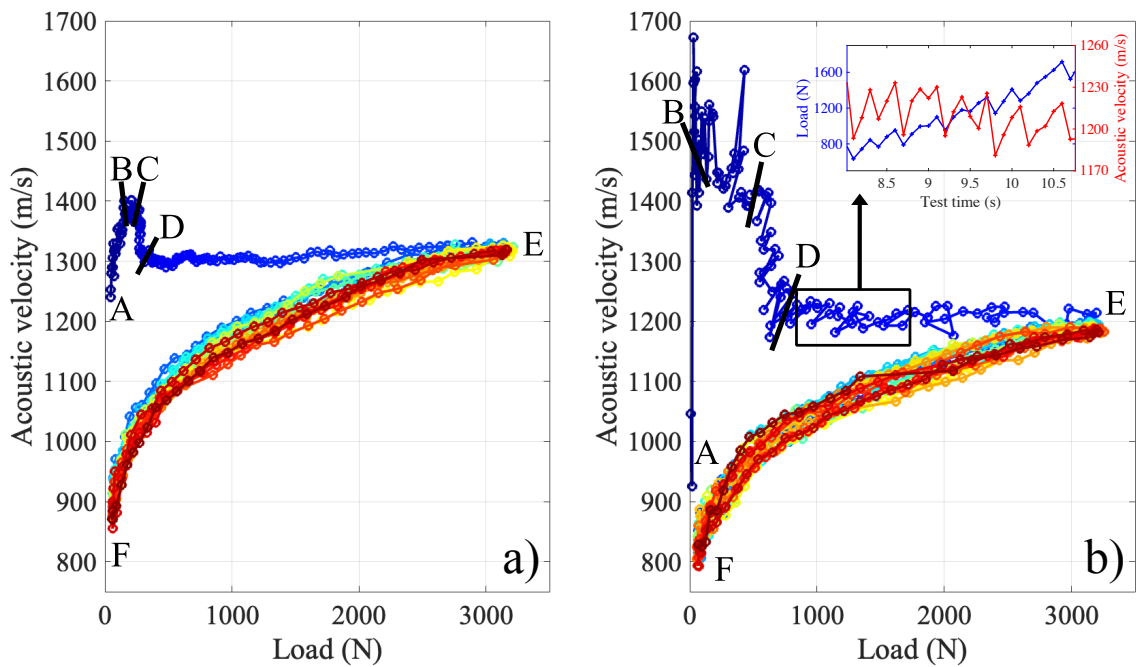

Figure 5. The velocity of the compressional coherent wave measured as a function of the load (strain driven) in the ductile- (a) and brittle- (b) cement bonded samples. The first loading is in dark blue, the last cycle of unloading-loading is in red, other colors refer to the intermediary cycles. Inset in (b) shows the correlation between stress drops and P-wave velocity decreases.

However, some significant differences on the evolution of $V_P$ are observed between these two cemented samples. In the case of the brittle cement, $V_P$ reaches a maximum value of about 1550 m/s when loaded from A to B, which is higher than in the case of ductile cement $V_P \approx 1400$ m/s. This difference stems likely from the higher stiffness of the brittle cement (salol) which can also be seen from the mechanical responses during the first loading: for example, at the load of 500 N, the deformation is about two times larger in Fig. 3(a) than that in Fig. 3(b). Nevertheless, $V_P$ collapses abruptly to a low value of 1200 m/s (a decrease of 23%) when the confining pressure is beyond a critical value (around point C) due to the brittle-like breakage of bonds. Instead, $V_P$ only decreases by 7% (down to 1300 m/s) under inelastic compaction in the ductile-cement bonded sample, pointing to a partial damage or breakage of bonds. Moreover, we find that for the brittle-cement bonded sample the fluctuation of the P-wave velocity is correlated to the intermittent stress drops as shown in the inset of Fig. 5(b) where the decrease of $V_P$ coincides almost with the stress drop. As mentioned above, the formation of shear bands is unlikely in the oedometric loading, thus the intermittent stress drops accompanied by the simultaneous weakening of the P-wave velocity could be associated with the formation and evolution of compaction bands [13]. This is reminiscent of a brittle-like failure or stress drop observed during shear banding or localization [10], in which an avalanche or cascade of local ruptures occurs through the correlated elastic interaction, accompanied by acoustic emissions (see below).

*Acoustic emission detection:* The *passive* probe consists in recording with an acoustic transducer the acoustic emissions (AE) due to the irreversible events occurring inside the sample under external loading [8, 20, 25]. They are equivalent to seismicity in the earth crust [26]. Monitoring the AE allows probing the precursor events such as cracks nucleation on the microscopic (local) scale before the material failure, i.e., fractures on the macroscopic scale. However, unlike previous works (e.g., [13]), 3D locations of AE are not possible in this work, due to the multiple wave scattering associated with the strong heterogeneity in our cemented granular samples [17].

Fig. 6 illustrates an example of AE recorded in a brittle-cement bonded granular sample, loaded at a speed of 50 μm/s. The sampling frequencies for the mechanical and acoustic measurements are 2.5 Hz and 500 kHz, respectively. During the first monotonic loading, intermittent stress drops occurrence increases with increasing the load. They are correlated to the increasing AE activity likely due to the bond damage or/and rearrangement as observed in dry granular media [20] under oedometric compression. These characteristic stick-slip like stress drops are however significantly reduced (or mostly disappear) in the subsequent cyclic unloading and reloading, and accordingly the medium becomes almost aseismic (or silent). Nevertheless, there is a non-stationary background acoustic signal, identified as the press-induced noise increasing with the load. It has been also observed with the press in the absence of the granular sample but just loading the piezoelectric (ultrasonic) transducers. This instrumental noise is strewn with intermittent AE of various amplitudes and spectral contents. In the ductile-cement bonded samples, AE are also recorded even with much less events and without visible stress drops.

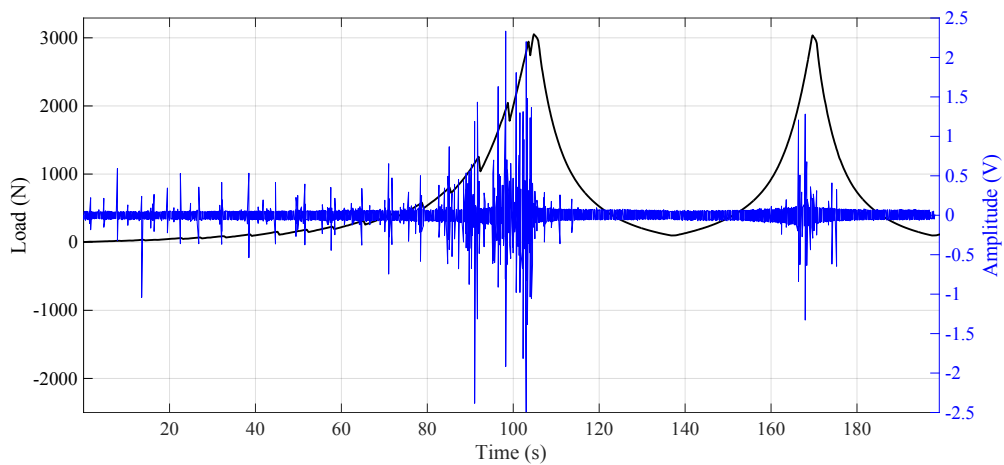

Figure 6. Load (black) and AE (blue) measured as a function of the time in a brittle-cement bonded granular sample during oedometric loading with a controlled axial displacement. Stick-slip-like stress drops are accompanied with significant AE.

To detect and analyze correctly AE in the continuous set of data with an undesirable noise, several classical methods exist: with a simple amplitude threshold or with the ratio of a short-term average on a long-term average (STA/LTA). Among numerous methods proposed including machine learning, we choose to use a simple k-means clustering spectral method based on the spectrogram of the continuous signal [27, 28].

We may characterize observed AE as *labquakes* and evaluate the maximum of the absolute amplitude $A_{AE}$ of each AE. The relative magnitude of each event is calculated according to $M = \log(A_{AE}/\min(A_{AE}))$. The probability distribution is plotted in Fig. 7, which may be compared to the Gutenberg-Richter law in seismology. It predicts that for a sufficiently large number of events $\log[N(m > M)] = a - bM$ with $N(m > M)$ the number of events whose magnitude $m$ is greater than $M$. Here the $b$-value is evaluated with a fit in the range of magnitude which we assume there are enough detected AE events. Indeed, the number of is underestimated both for lower magnitude due to the detection sensitivity and for ide events which are rare. For the earthquake statistics, the expected standard $b$- ir estimated value of $b \approx 1.3$ is consistent with other similar experimental studies characteristic stick-slip events give rise to a hump tail (with a local maximum) at gnitude of about $M = 1.6$.

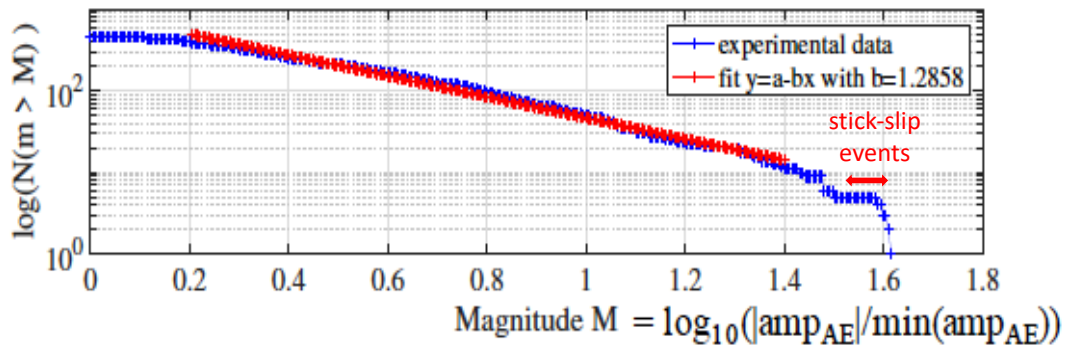

ability distribution of AE occurrence versus magnitude (in blue) plotted in a log-log scale eristic contribution due to quasi-periodic stick-slip stress drops (indicated by the double fit is based on the Gutenberg-Richter law (in red).

1age evolution under cyclic loading with increasing consolidation stress anical response: To better understand the damage process during inelastic in cemented granular materials, we complete here the above experimental with increasing consolidation stress under cyclic loading. As shown in insets of nvestigate the response of the sample to cumulative damages by increasing the maximum force (i.e., consolidation stress) of cyclic loading, unloading and m $F = 500$ N to 3000 N by a step of 500 N while the minimal load is kept constant ve investigation. The sampling frequencies for mechanical measurements are 20 for tests in ductile- and brittle- cement bonded samples, respectively. For each loading cycle, the displacement varies linearly with time at a speed of 50 μm/s which is about two time slower than the loading rate in Fig. 3, except immediate inversions of the loading direction displacement due to the inertia of the electromechanical press that induces a short relaxation of about 0.6 s.

Fig. 8 depicts the cross-plots of loading force versus displacement in the case of the ductile (a) and brittle (b) cements. They highlight the anelastic and irreversible behavior under cyclic unloading and reloading, already observed in Fig. 3. The unloading curve presents systematically a different slope from the preceding loading path with increasingly steep slopes. As long as the load remains lower than the previous consolidation load, a slight hysteresis between unloading and reloading is observed at each cycle. Nevertheless, if the reloading exceeds the consolidation load (end point), they recover the master-like curves which are resembling the force-displacement responses under the monotonic loading in Fig. 3 (the vertical drop at the beginning of the unloading corresponds to the displacement inertia of the press and the induced relaxation). Note however these master curves obtained at a lower loading speed (i.e., 50 μm/s instead of 100 μm/s) show weakened rigidities compared to those observed in Fig. 3. This may be due to the visco-plastic behavior of cements (lower the load speed is, higher the relaxation of deformation is) and also to the more pronounced damage associated with the ramped cyclic consolidation stress, in particular for the case of the brittle cement for which the compressive displacement is about 3.7 mm in Fig. 8(b) but 1.2 mm in Fig. 3(b) when the load reaches $F = 3000$ N. At the highest loading observed here, our cemented samples appear to be totally damaged, namely, nearly no more breakage of bonds are possible on further increasing the load. As a result, the master curves and the reloading paths merge with the same slope, corresponding to the usual behavior of a dry granular material under cyclic loading-unloading (i.e., states between E and F in Fig. 3).

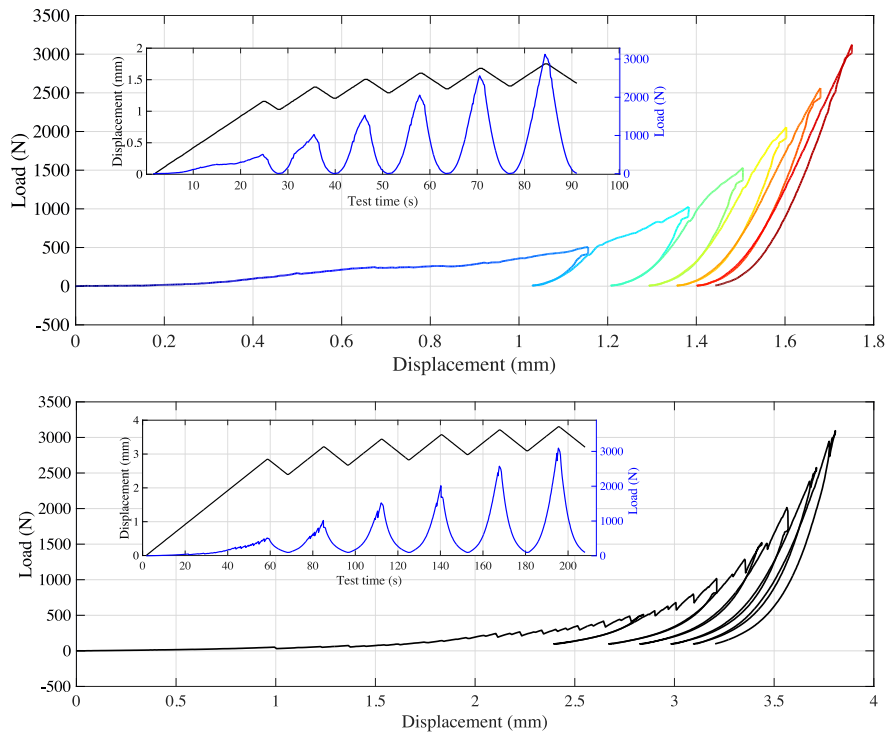

Figure 8. Mechanical responses under oedometric cyclic loading with increasing consolidation stress of the ductile- (a) and brittle- (b) cement bonded granular samples. Insets show loading forces and displacements measured as functions of test time.

Moreover, in the case of the brittle-cement bonded sample (Fig. 8(b)), the stress drops (indicated by saw-tooth-like stick-slip oscillations) are absent all along the unloading and during the reloading when the load is lower than the maximal load reached previously (also called consolidation stress). However, they reappear when unexplored greater values of the load are reached as if the sample has a memory of the loading history. This idea of memory and new states reached when the load increases will be confirmed below by the AE analysis and Kaiser effect. The absence of apparent hysteretic loops suggests a negligible frictional dissipation here.

*Ultrasound velocity measurement*: As in section 2.2.2, we can also investigate the material structural change during inelastic compaction loading by monitoring the evolution of the P-wave velocity. Fig. 9 depicts the P-wave velocity measured as a function of the loading force (or axial stress) with increasing the consolidation stress in the ductile cement-bonded granular samples. Compared to the data obtained with the constant consolidation stress protocol (Fig. 5(a)), we observe here an overall decrease of the P-wave velocity at a given load, with the increase of the consolidation stress of the cyclic loading. We note narrow hysteretic loops. Such hysteretic loops of P-wave velocity recover finally those obtained with constant consolidation stress at the same load range, as shown by the last cycle very similar to that in Fig. 5(a) (red curves). The main difference between the two loading protocols lies in the P-wave velocity $V_{bond}$ (see section 3.2), measured upon unloading at (almost) *vanishing* pressure $P \approx 75$ kPa ($F \approx 60$ N). Here $V_{bond}$ decreases progressively with repeated unloading-reloading cycles from $V_{bond} \approx 900$ m/s to 750 m/s ($\Delta V_{bond}/V_{bond} \sim 17\%$), while it remains at relatively higher value and varies little from 940 m/s to 860 m/s ($\Delta V_{bond}/V_{bond} \sim 8.5\%$) in Fig. 5(a). This observation suggests that oscillatory (cyclic) loading, may facilitate the damage process via the progressive bond breakage (see discussion below), reminiscent of a fatigue process.

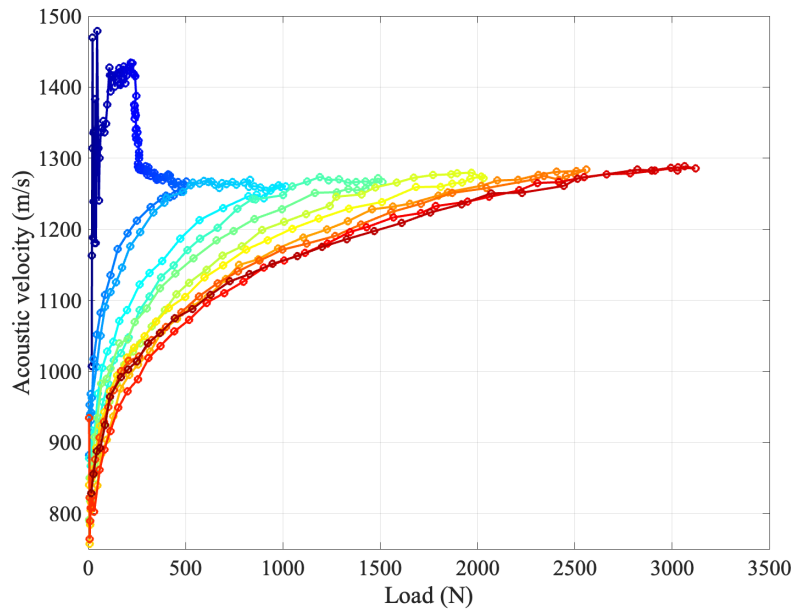

Figure 9. P-wave velocity measured with the time of flight of the direct P-wave as a function of the loading force in a ductile-cement bonded sample.

*Acoustic emission detection:* As a passive acoustic probing, we also show in Fig. 10 the acoustic emission recorded in the brittle-cement bonded sample under increasing consolidation stress. It is sampled at 200 kHz continuously all along the mechanical test lasting for about 3 minutes. The noise of the electromechanical press is clearly visible as its amplitude evolves with the imposed load, as in the cyclic loading test with a constant consolidation stress. As seismicity, these AE are strongly correlated to the intermittent stress drops (Fig. 10(a)). We will analyze these AE in more details below (section 3.3), in particular with the memory effect observed above in section 2.3.1.

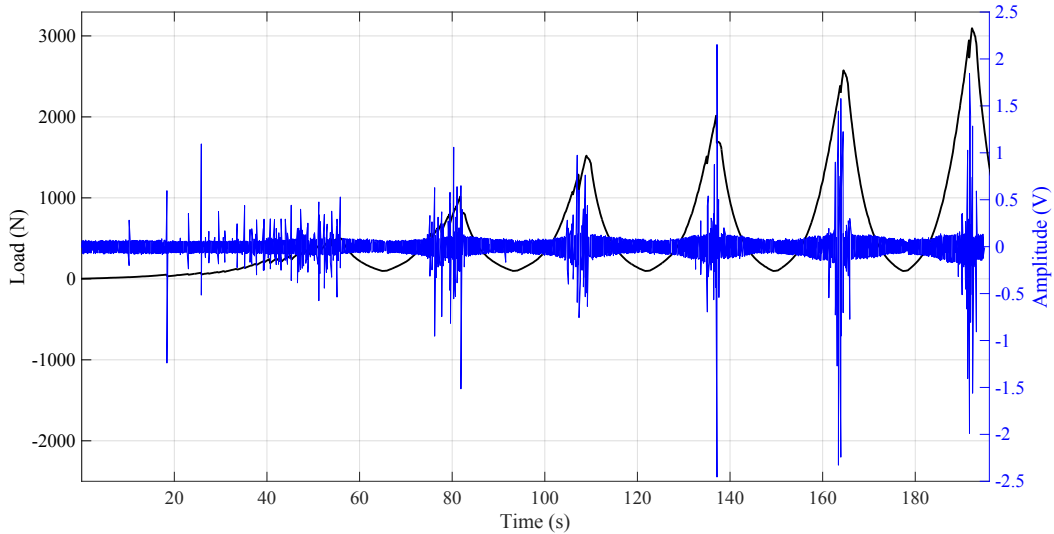

Figure 10. Measured loading force (black) and AE (blue) as functions of test time in a brittle-cement bonded sample, under cyclic loading with increasing consolidation stress.

## III. DISCUSSION AND MODELING

Inelastic compactions have been observed under oedometric loading beyond certain threshold, monotonic (Fig. 3) and cyclic with increasing consolidation stress (Fig. 8), in cemented granular materials. The induced damage and plastic deformation investigated here are associated both with the breakage of cement (ductile and brittle) bonding and the rearrangement of grains, without the grain crushing. In opposite, it has been shown that inelastic compaction in porous sandstone during triaxial experiment or hydrostatic compaction (zero deviatoric stress) is associated with grain crushing and pore collapse [14, 15 30]. They observed two possible failure modes: localized compaction bands at low confinement and more complex diffuse compactions at higher confinement, accompanied by abundant acoustic emissions events.

Our experimental observation of simultaneous intermittent stress drops, the P-wave velocity decreases and AE detection in the brittle-cement bonded sample indicate that the inelastic compaction observed here are dominated by the material *bulk* structural change and damage, not by the sliding (stick-slip) of the sample at the interface with the oedometer cell. To monitor such a material damage, both active and passive acoustic detections provide indeed adequate and unique probing.

**A. Interplay between bond damage and contact hardening under compression**

Physically, two distinct and competitive mechanisms are affecting the elastic wave velocities during inelastic compaction: debonding (cracking) and hertzian contact (porosity decrease). We seek to interpret the evolution of P-wave velocity observed (Fig. 5) within the effective medium framework where small amounts of cements are assumed to be distributed homogeneously between grain contacts, forming independent bonds in the (so-called) pendular regime (Fig. 11(a)) [18].

On the scale of cemented grain contact (Fig. 11(b)), we consider a heuristic model proposed by Dvorkin et al [31, 32] with a parallel association of two mechanical elements: the first is a nonlinear spring $k_H$ for describing the hertzian-like contact between beads, and the second corresponds to an elasto-plastic element $k_B$ for modelling the cemented bond. The contact stiffness $k_H(P)$ increases with increasing the (isotropic) compression $P$, while $k_B$ is a constant (linear spring) that breaks down irreversibly to zero (healing effect is not considered here) beyond the yield threshold. Therefore, the stiffness of a cemented contact resulting from the combination of these two parallel elements, $k_C \sim k_H + k_B$, may increase with increasing the compression $P$, then saturate or decrease due to the evolutions of $k_H$ and $k_B$ in counteracting dependence with $P$.

Within the effective medium theory (EMT), the compression and shear wave velocities are $V_P = [(K + 4G/3)/\rho]^{1/2}$ and $V_S = (G/\rho)^{1/2}$ where the bulk ($K$) and shear ($G$) effective modulus are linearly proportional to the microscopic contact stiffness $k_C(P)$ via the coordination number $Z$ and implicitly the porosity $(1 - \phi_s)$ (see Eq. (2) below). Consequently, the evolution of the P-wave velocity in Fig. 5 may be correlated to the change of the contact stiffness $k_C(P)$ and interpreted as follows. At first, increasing the oedometric loading $P$ makes $V_p$ increase due to the increase of $k_H$ and also the possible creation of new contacts affecting the coordination number $Z$ [33] where the cemented bonds behave elastically [17, 32] with a constant $k_B$ (from A to B). When the damage process (debonding or crack nucleation) becomes dominant, $V_p$ first decreases slightly (from B to C) and then drastically with likely the formation of compaction bands, i.e., macro-fractures within a set of intact beads ('plateau' from C to D), as those occurred in porous sandstones with pore collapse (and also grain crushing). Further increasing the load, $V_P$ reaches asymptotically to a non-cohesive granular state [21] where most of cemented bonds are broken or damaged nearly, as it can be expected.

Indeed, upon unloading and reloading (between E and F), we recover the similar evolution of P-wave velocity $V_P$ versus the confining stress $P$, as those observed in non-cohesive (dry) granular materials $V_P \sim P^\beta$ with $\beta$ between 1/4 and 1/6 [24, 33, 34]. Note however there still exists a residual bonding (Langlois and Jia, 2014), since the P-wave velocity $V_{bond}$ remains important (Fig. 5) upon unloading at $P \approx 0$ (see section 3.2). The above situation is different with those observed in Fig. 9. When unloading-reloading is performed at a low consolidation stress (i.e., blue curves), for which a large number of cemented bonds is not yet damaged, $V_p$ does not reach the non-cohesive granular states obtained between E and F (red curves) in Fig. 5(a). Instead, $V_p$ at the "end points" (i.e., when the unloading-reloading is finished and the reloading reaches the previously applied consolidation stress) follows the path D-E (blue curve). This observation shows that the partially damaged sample has a memory of its prestressed state.

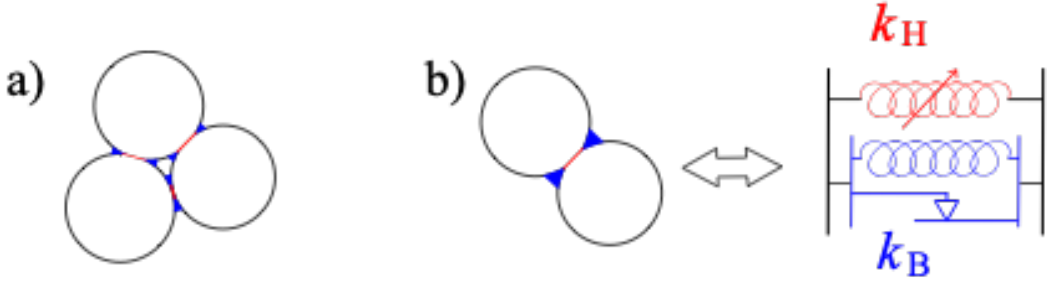

Figure 11. Schematic illustrations of cemented granular materials: (a) cement bonds are independent (not link between them), i.e., in the pendular regime; (b) cemented contact is modelled as a parallel association of a hertzian contact and an elasto-plastic element.

**B. Evolution of residual cohesive bonding under cyclic loading: Digby's model**

As mentioned above, to investigate the damage or debonding degree by the oedometric loading, we may evaluate the P-wave velocity $V_{bond}$ at $P \sim 0$. To support this idea, we consider the bonded contact model proposed by Digby [35]. As illustrated in Fig. 12, the contact area between two elastic spheres (of radius $a$) includes a bonded portion (of radius $b$) to simulate a cementing or sintering effects between grains [36]. When increasing the pressure (which is approximated as the axial stress $P$), the total area of contact increases to radius $a$ ($\geq b$). The radius $b$ is independent of the applied compression; $b = 0$ corresponds to the case of non-cohesive particles. By analogy with the Hertz theory and the Mindlin model, the normal and tangential contact stiffnesses, $k_n$ and $k_t$ respectively, are given by

$$k_n = 4\mu a/(1-v) \quad (1a)$$

$$k_t = 8\mu b/(2-v) \quad (1b)$$

if the annular contact region between radii $a$ and $b$ is assumed to be perfectly sliding. Here $v$ and $\mu$ are the Poisson ratio and the shear modulus of the grain material. By a statistical analysis (i.e., effective medium approach), the effective bulk $K$ and shear modulus $G$ can be derived from the contact stiffnesses $k_n$ and $k_t$ as $K = Z\phi_s k_n/(12\pi R)$ and $G = Z\phi_s(k_n + 3k_t/2)/(20\pi R)$ with $R$ the bead radius [37]. From Eq. (1), the elastic moduli $K$ and $G$ and accordingly the elastic wave velocities $V_P$ and $V_S$ may also be given by the contact radii $a$ and $b$ [35],

$$V_P^2 = (\mu Z/5\pi R \varrho_g)[3a/(1-v) + 4b/(2-v)] \quad (2a)$$

$$V_S^2 = (\mu Z/5\pi R \varrho_g)[a/(1-v) + 3b/(2-v)] \quad (2b)$$

where $a$ depends on the applied pressure as $\sim P^{1/3}$ (hertzian contact) and $b$ is attributed to the bonded contact area. Together with the coordination number $Z$ and the bead property, this bonding (cohesion) parameter $b$ gives a wave velocity $V_{bond} \sim (Z b)^{1/2}$ at vanishing pressure, providing thus a measurable indicator of debonding (damage) by acoustic velocity, as mentioned above.

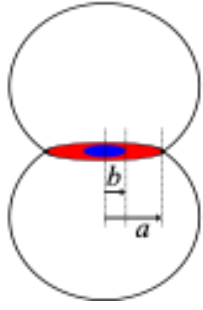

Figure 12. Digby's bonded contact model with a total circular contact area of radius *a* made of an annular sliding area (in red) and a bonded area (in blue) of radius *b*.

The analytical damage model of cemented granular materials (on the macroscopic scale) is not available despite several theoretical efforts (e.g., [31]) and numerical simulations [11, 12, 38, 39]. The simulation of oedometric compression experiments developed in Canel et al [39] point out that the strain field are very heterogeneous and that a small amount of damage of bonds on the local scale induces a dramatic decrease in the coherent wave velocity on the global scale, which provides thus a very sensitive probe to monitor the damage processes in such heterogeneous materials.

Nevertheless, the above EMT based on the Digby's bonded contact model was used to interpret empirically the evolution of elastic wave velocities measured in porous sandstones during inelastic compaction [30]. More specifically, the significant increase of $V_P/V_S$ ratio (which depends on the quantity $a/b$ according to Eq. 2) beyond a certain threshold stress can be related to the decrease of *b*, i.e., the bonding radius. This implies that the grains became less and less cemented with increasing the external loading, which was consistent with their observed grain crushing.

To investigate furthermore the debonding (damage) in this work, we identify as a first approximation $k_H$ and the elasto-plastic element $k_B$ in Fig. 11(b) as $k_n$ and $k_t$ in Eq. 1, respectively. In order to fit the measured P-wave velocities in cemented samples (Fig. 5), we rewrite the P-wave velocity as,

$$V_P^2 = \alpha F^{2\beta} + V_{bond}^2 \quad (3)$$

Here the first term originates from the Hertz contact law with *F* the axial loading force and $\alpha$ a constant whereas the second term refers to the cemented bond-induced velocity $V_{bond}$ (found at vanishing pressure) characterizing the degree of bond damage under loading, monitored at vanishing load (unloading) $F = 0$. Unlike previous works in porous sandstones, the damage process would be here dominated by the debonding of cements between grain contacts rather than grain crushing [13]. This should make the above Digby's model more relevant for the case considered in this work.

There are two main parameters $V_{bond}$ and $\beta$ to infer in Eq. (3) from P-wave velocities measured in the weakly cemented granular state, between E and F (Fig. 5). More specifically, we fit with the data obtained during unloading in order to monitor the damage degree by the previous loading. By choosing an exponent $\beta = 1/6$, we can deduce $V_{bond}$ as the only free parameter of a curve fitting procedure, with $\alpha$ found nearly constant for each unloading cycle. Fig. 13 shows $V_{bond}$ inferred from tens of cycles in the ductile cement-bonded sample with constant and increasing consolidation stresses, respectively. With constant consolidation stresses (Fig. 5(a)), the fitted $V_{bond}$ varies slightly with loading-unloading cycles around a value of 600 m/s, corresponding to $F = 0$ in Eq. (3) (except the discrepancy at the cycle # 6, likely due to a manipulation defect). They are lower than the previously estimated $V_{\text{bond }d}$ measured at the vanishing load $F \sim 60$ N. One should note that this fit procedure assumes $V_{\text{bond}}$ to be constant during a given unloading, however it may vary with new damages for instance.

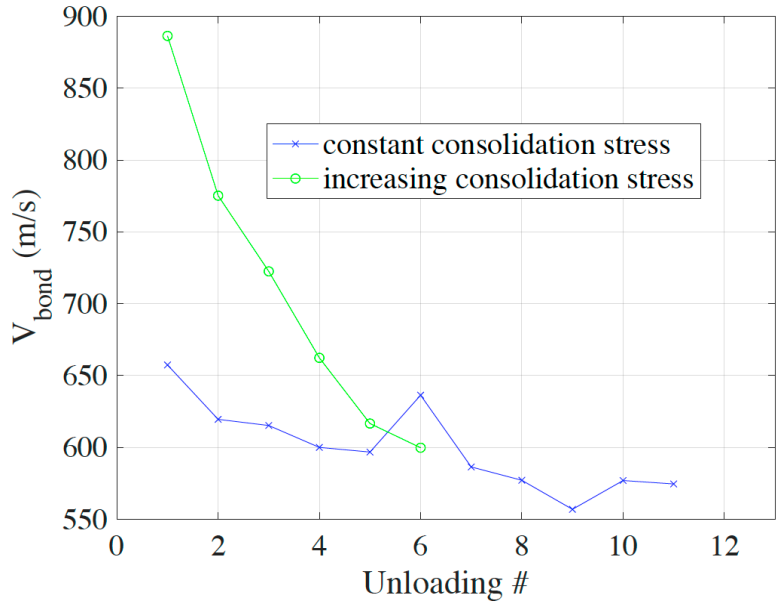

Figure 13. $V_{\text{bond}}$ measured as a function of loading cycles in the ductile-cement bonded sample, under cyclic loading with *constant* (blue crosses) and *increasing* (green dots) consolidation stresses, inferred from measured data $V_P(F)$ via Eq. (3) with $\beta = 1/6$.

We now examine the P-wave velocity evolution, still with the model of Eq. (3), under cyclic unloading-reloading with increasing consolidation stress and focus on the ductile-cement bonded sample (Fig. 9). Using the same fitting procedure described above, we infer the cohesion velocity $V_{\text{bond}}$. After the first unloading with $V_{\text{bond}}$ found close to 900 m/s, Fig. 13 illustrates a considerable decrease of $V_{\text{bond}}$ with successive loading cycles, from 860 m/s to 560 m/s. This acoustic probing confirms that additional damage (debonding) is created with increasing the consolidation stress as can be expected. Interestingly, it also reveals that for a given consolidation stress, the cyclic ramping protocol is more efficient than the monotonic loading to transition the cemented materials to non-cohesive granular packings. The structural relaxation upon unloading may facilitate the diffusion of cracks (debonding), e.g., via rearrangement of grains [22], which enhances the fracture nucleation upon reloading, reminiscent of a fatigue-like process.

### C. Acoustic emissions during inelastic compaction: Omori-like law and Kaiser effect

In this study, AE (i.e., seismicity) observed in brittle-cement bonded granular samples are strongly correlated with recorded stick-slip-like stress drops. Investigation of the cumulated number of such AE and the rates (number of AE per second) is of considerable importance for understanding precursor events. Fig. 14(a) illustrates the accumulated number of events measured under cyclic loading with constant consolidation stress (Fig. 6). It increases drastically during the *stick* phase up to a large stress drop, i.e., the 'mainshock'. The rate of these AE (precursors) is influenced by both the loading stress $P$ or force $F$ and the proximity $F$-$F_{drop}$ to the stress drop (failure) characterized by the sudden decrease $\Delta F_{drop}$.

We here focus on the evolution of the (normalized) AE rate during the stick phase for the ten stick-slip events, from #5 to #14 (Fig. 14(b)). The events from #1 to #4 are not considered because the stress drops and consequently AE are too weak to be detected efficiently through the k-means clustering method. The last event #15 is also neglected here due to the measurement incertitude affected by the sampling frequency of our electromechanical press (2.5 Hz) which is likely too low to measure correctly the loading stress change during this abrupt stress drop event. These AE rates are calculated over a moving window of 3 s with an overlapping of 97%, which makes it continuously decrease at the end of each stick phase by averaging effect with the rates of the following consecutive phase.

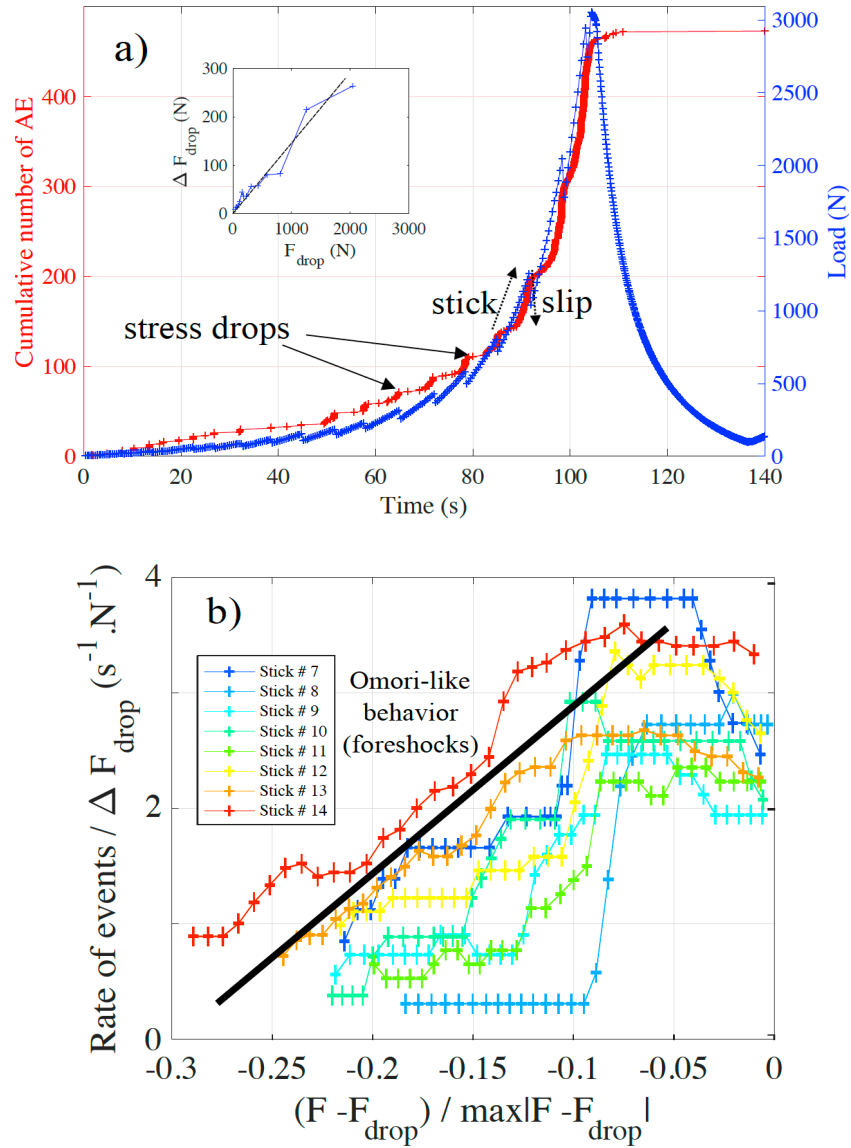

Figure 14. (a) Load (in blue) and cumulated number of acoustic emissions (in red) as functions of the time in a brittle-cement bonded sample. Inset: the magnitude of stress drops as a function of the threshold stress (failure). The dashed line indicates an effective friction law with a friction coefficient $\mu \approx 0.15$. (b) Temporal events rates during the stick phases #7 to 14 normalized by the corresponding stress drop magnitude as functions of the normalized distance of load to failure, i.e., large stress drop (see text). The black curve suggests an Omori-like law for foreshocks.

Remember that the magnitude of *shear* stress drops $\Delta\tau$ associated with the stick-slip instability in sheared granular layers remains comparable at the constant normal stress $P$ (Marone, 1998). Interestingly, we observe here an almost linear increase of the stress (force) drop $\Delta F_{drop}$ with increasing axial load $F$ during inelastic compaction shown in the inset of Fig. 14(a). This is indeed formally reminding of a form of friction process. If we assume $\Delta\tau \sim (\mu_s - \mu_d) P$ (with $\mu_s$ and $\mu_d$ the Coulomb-like static and dynamic friction coefficients), by analogy $\Delta\tau$ versus $\Delta F_{drop}$ and $P$ versus $F$, we may speculate an increase of $\Delta F_{drop}$ with increasing $F$ assuming a friction coefficient about 0.15. On the basis of this analogy, we plot the normalized AE rates by $\Delta F_{drop}$ as a function of the distance to the stress drop $(F - F_{drop})$ normalized by max $(|F - F_{drop}|)$ for the series of stress drops. Fig. 14 (b) shows such kind of plots where all AE data seem to gather around a unique (master) curve, suggesting thus a scaling law which may be reminiscent of Omori-like law for foreshocks [20]. Considering the difference of loading protocols between direct-shear and oedometric testing, further investigation is needed to better understand the quasi-regular stick-slip behavior associated likely with inelastic compaction bands.

Fig. 15(a) shows the AE recorded during cyclic loading with increasing consolidation stress (as in Fig. 8) and the cumulative number detected with a STA/LTA process: 0.1 ms for the STA, 5 ms for the LTA and a detection threshold of 0.1 [27]. The cumulative number of 1295 AE recorded as a function of the loading history (Fig. 15(b)), i.e., loading, unloading and reloading, confirm more precisely the material memory property seen in mechanical (Fig. 8) and P-wave velocity measurements (Fig. 9) for the two first cyclic loadings where the AE rates recover the master curve at end-points upon reloading. Such observations, known as the Kaiser effect, exist widely in amorphous solids such as metal alloys [40] and rocks [41-43]. AEs are generally linked to the creation and development of fractures in a solid-like material under loading. It has been shown by the localization of AE sources that during unloading there are nearly no AE due to the absence of the crack growth, but during reloading AE occur in the previously damaged zones with the growth of fractures, particularly when the load becomes greater than the previous consolidation stress [44].

However, we here observe that the Kaiser effect is less and less respected when the load is beyond certain threshold ($F \sim 900$ N), meaning that the sample is so damaged that it starts losing its memory property. The cemented granular material undergoes thus a transition from cohesive (solidlike) to granular-like (fluidlike) states where the stress principal axes may evolve with the loading history [45-47]. Such an evolution is also indicated by the decrease of P-wave velocity $V_{bond}$ at vanishing pressure (Fig. 13(b)).

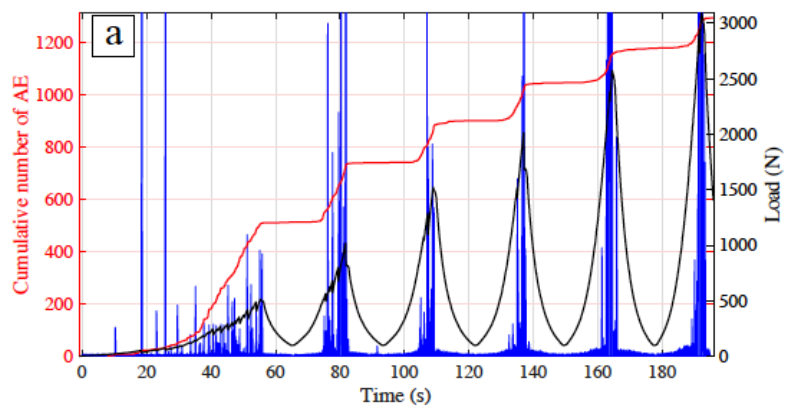

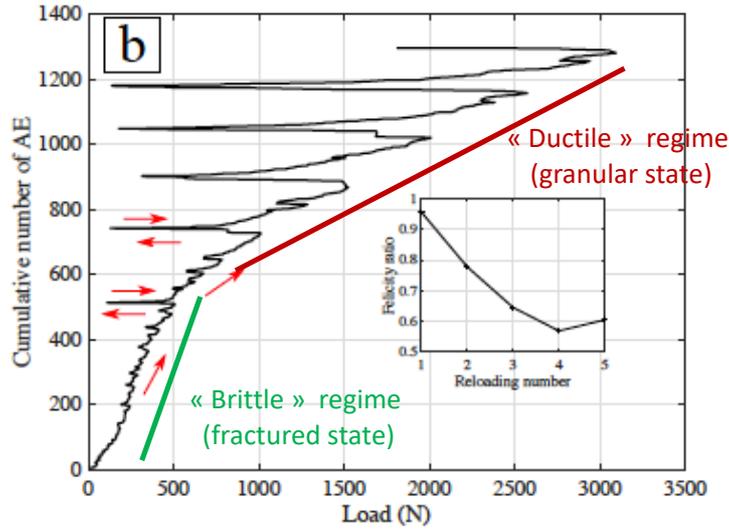

Figure 15. (a) Load (black curve), absolute value of the AE recording (blue curve) and cumulative number of AE (red curve) as functions of the time. b) Cumulative number of AE as a function of the load. Red arrows indicate the followed path in time until the second reloading. The inset shows the five Felicity ratios available from this test.

To characterize the occurrence of AEs upon reloading at a lower load (already explored) than the previous consolidation stress (maximal), we may evaluate the Felicity ratio which is considered as measure of the (rock) material quality [42]. It is defined as the ratio between the applied load at which new AE occur during the reloading and the previous maximum applied load (consolidation stress). In this work, we consider arbitrarily that new AE appearance becomes significant from ten more AE upon the reloading for calculating the Felicity ratio. This ratio is equal to 1 for the perfect Kaiser effect (in practice, we consider that ratio for a value greater than 0.9). The five Felicity ratios available from this test are plotted in the inset of Fig. 15(b). The first cycle highlights a nearly perfect Kaiser effect, i.e., the memory effect, as observed above in mechanical (Fig. 8) and acoustic P-wave velocity measurements (Fig. 9), with the Felicity ratio equal to 0.96. It then decreases down to 0.57 at the fourth reloading before increasing slightly to 0.61 at the last reloading.

Finally, we remind that the Kaiser and Felicity ratios are usually investigated with triaxial tests and not with oedometric loading. The main difference is that in triaxial tests, compression strength is well defined corresponding to the maximal load reached before the load decrease, whereas in oedometric tests, such a compression strength is not clearly observed. Nevertheless, our work suggests that the Kaiser and Felicity effects observed here could correspond to the compaction banding, localized or diffuse, similar to those in porous sandstones [14, 15, 30] but without grain crushing observed in our works. Such evaluations provide also a possible measure of the material damage, i.e., the distance to failure. However, it is still not clear why no significant AE are detected in our ductile-cement bonded samples (aseismic) during inelastic compaction.

## IV. CONCLUSION

In summary, we have investigated the inelastic compaction of cemented granular materials under the oedometric loading, monitored simultaneously by measuring the P-wave velocity (active detection) and recording the acoustic emission if any (passive detection). The nonlinear mechanical behavior with an increasing loading is related to the transition from a cohesive to a partially non-cohesive granular material via the debonding of cemented grain contacts. This damage process is manifested clearly by the elastic weakening on the macroscopic scale through the decrease of the P-wave velocity. Using the Digby's cemented granular model, we can explain correctly our P-wave velocity measurements and make the cohesion velocity $V_{bond}$, measured at vanishing pressure, as an indicator of the debonding on the scale of grain contact. A main finding is that as a function of the bond material, ductile or brittle, a cemented granular sample behaves fundamentally differently. In the former the transition during inelastic compaction undergoes continuously suggesting diffuse compaction bands, while in the latter the cohesive to non-cohesive state transition via plastic deformation is accompanied by intermittent stress drops (stick-slip) pointing to the localized compactive shear bands on the mesoscopic scale. These observations may be understood by a short-range interaction for the ductile-cement bonds (due to viscous damping) but a long-range interaction for the brittle-cement bonds (via elastic response).

The fundamental difference between these two types of cemented granular samples is also revealed by their activity of acoustic emission: the former is nearly aseismic while the latter has as microseismic activity. AE characterize the damage process and microscopic physics on the grain contact scale. The aseismicity in ductile-cement bonded samples may be explained by two effects: the first is the overdamped motion of grains after the bond breakage (AE source) due to the viscous ductile cement and the second is the high attenuation of emitted wave propagation (if any) through the contact network caused by these dissipative bonds [48]. Instead, AE observed in brittle-cement bonded samples demonstrate clearly their correlation with the stick-slip-like stress drops, as observed in sheared granular layers and seismic faults. These AE show a statistical distribution in agreement with the Gutenberg-Richter law demonstrating thus the powerful similarity between these labquakes and seismicity [26]. They also indicate a drastic increase of AE rates when the load approaches a failure, i.e., a large stress-drop event or mainshock, the temporal distribution being close to the linear Omori-like law for foreshocks. With the consolidation stress ramped up during the cyclic loading, AE rates slow down together with the loss of the (Kaiser) memory effect. This occurs precisely at the transition of a cemented porous material to a non-cohesive granular medium through fracture nucleation process [21]

Surprisingly, quasi-periodic large stress-drops akin to stick-slip events in granular faults under shear have been observed here during the oedometric loading. Contrary to what happens in non-cohesive granular materials, these stick-slip instabilities may be associated with the formation of compaction bands due to the cohesion-induced long-range interaction, resulting in a brittle-like fracture nucleation [13, 15]. Further investigation is needed to better understand the underlying physics in the presence of shear banding, which is also of interest for the earthquake rupture [49]. Active monitoring by multiple scattered ultrasounds and DEM simulations on the grain-contact scale [24, 50] may help for such a study.

**ACNNOWLEDGMENTS**

The authors acknowledge support from the European Research Council under the European Union Horizon 2020 research and innovation program (grant agreement no. 742335, F-IMAGE) and also from PSL University under the program "Investissements d'Avenir" launched by the French Government.